\documentclass[12pt]{article}
\usepackage{amsfonts,epsfig}
\usepackage{amsmath,amsfonts}
\usepackage{amssymb,graphicx}

\def\pa{\partial}

\def\d{\delta}
\def\h{\hat}
\def\b{\bar}
\def\t{\tilde}
\def\f{\frac}

\def\p{\varphi}

\def\d{\delta}
\def\h{\hat}
\def\t{\tilde}
\def\f{\frac}

\def\l{\label}
\def\e{\varepsilon}
\def\a{\alpha}
\def\bt{\beta}
\def\la{\lambda}
\def\g{\gamma}
\def\G{\Gamma}
\def\m{\mu}
\def\n{\nu}
\def\w{\wedge}

\def\r{\rho}
\def\s{\sigma}
\def\S{\Sigma}
\def\th{\theta}

\def\be{\begin{equation}}
\def\ee{\end{equation}}
\def\ba{\begin{eqnarray}}
\def\ea{\end{eqnarray}}
\def\z{\bar{z}}

\def\r{\rho}
\def\s{\sigma}
\def\S{\Sigma}
\def\th{\theta}

\def\be{\begin{equation}}
\def\ee{\end{equation}}
\def\ba{\begin{eqnarray}}
\def\ea{\end{eqnarray}}
\def\z{\bar{z}}
\def\ea{\end{eqnarray}}
\def\z{\bar{z}}
\def\c{\chi}
\def\x{\tilde{x}}
\begin{document}

\vspace{5cm}
\centerline{\large\bf Surface charge of horizon symmetries of
a black hole with supertranslation field}
\vspace{1cm}
\centerline{{\bf Mikhail Z. Iofa}
\footnote {e-mail iofa@theory.sinp.msu.ru} }
\centerline{Skobeltsyn Institute of Nuclear Physics}
\centerline{Lomonosov Moscow State University}
\centerline{Moscow, 119991, Russia}
\vspace{1cm}

\begin{abstract}

Near-horizon symmetries are studied for static black hole solutions to Einstein
equations containing  supertranslation field.
We consider general diffeomorphisms 
which preserve the gauge and 
 near-horizon structure of the metric. 
Diffeomorphisms are generated by the vector fields and form a 
group of near-horizon symmetries.
  The densities of variation  of the surface charge associated to asymptotic horizon
symmetries of the metric are calculated in different coordinate systems 
connected by "large" transformations which change supertranslation field
in  metric. 
Integrability of the variations of surface
charge is considered in different coordinate systems.
The case of integrable variation of the surface charge is studied in detail.
\end{abstract}
\vspace{3cm}

\section{Introduction}

The final state of gravitational collapse is a stationary
metric diffeomorphic to the metric of the Kerr black hole 
\cite{carter,robinson,chand}.
General diffeomorphisms contain  pure gauge transformations which are
changes of coordinates and "large" transformations which 
change  supertranslation field in 
metric. Physically "large" transformations map a
physical state to another  physical state with
a different cloud of soft particles \cite{fad,carney,strom4}.

Supertranslations naturally appear in a study of symmetries of the
asymptotically flat of gravity at the null infinity initiated by
Bondi, van
der Burg, Metzner and Sachs \cite{bondi,sachs}. The infinite-dimensional group
of the asymptotic symmetries (BMS group) extends the Poincare group and contains a
normal subgroup of supertranslations which are the angle-dependent
translations of retarded time at the null infinity \cite{strom1}.

 BMS algebra can be further enhanced to contain superrotations
\cite{bar2,bar3,bar4,charge}. Exponentiation of the infinitesimal 
supertranslation and superrotation
generators produces finite transformations, but in distinction to
supertranslations,  exponentiation of the infinitesimal
superrotations when acting on physical states leads to states with the
energy unbounded from below \cite{comp1}.
One cannot introduce a physical state with a finite superrotation
charge, but there exist  conserved charges associated with
supertranslations and superrotations \cite{bar2,bar3,bar4,charge}. 
Supertranslation charges
vanish except for a charge corresponding to the mass of a state, but
 finite superrotation charges differ for states with
different supertranslation fields.

   BMS transformations are naturally formulated at the null
infinity, but there is a complicated problem of extension of an
asymptotically defined metric containing a supertranslation field in
the bulk.
In paper \cite{comp1} a family of 4D
vacua containing a supertranslation field was constructed in the bulk,
and  in paper
\cite{comp2} a solution-generation technique was developed and the
black hole metrics diffeomorphic to the Schwarzschild metric and
containing a supertranslation field were obtained.

In this paper we study the near-horizon symmetries of the black holes
containing a supertranslation field. The near-horizon symmetries are
the main characteristics of horizon microstates which in turn define
 thermodynamic properties, the entropy and evaporation of a black hole.
Near-horizon symmetries were  extensively investigated in
a large number of
papers (very incomplete list of refs. is \cite{hen,car1,car2,
strom3,LW,IW,BB,WZ,hotta,koga,kang,palla,padm,donnay41,donnay4,akh,set1,set2,
maj,car3,haj1,mao} )

The near-horizon region is foliated by a set of surfaces enclosing the
horizon surface and located at a distance $x$ from the horizon
($x$ is defined differently in different coordinate systems).
Near-horizon symmetries are generated by transformations which  
preserve the horizon structure of a
metric, and  do not change the power of the leading in $x$
terms in the components of the metric considered at a near-horizon surface at a
distance $x$ from the horizon, as $x\rightarrow 0$.

We consider the near-horizon transformations of the static black hole 
solutions of the Einstein equations containing a supertranslation field. 
Transformations are 
generated by the vector fields $\xi^{\mu}$. 
The metric variations $\d_{\xi}g$ 
are elements of the tangent space to the space  of metrics and are
solutions to the linearized Einstein equations. On the tangent space is defined
a bilinear presymplectic form. The presymplectic    Lee-Wald form \cite{LW}  is
$$ 
w^ {\mu\,LW}(\d_1 g, \d_2 g, g)=\d_1\Theta^\mu (\d_2 g, g)
-\d_2\Theta^{\mu} (\d_1 g, g) -\Theta^\mu (g,\d_{[1,2]}g)     ,$$
where $\Theta^\mu$ is the boundary term in a variation of the Einstein action
$$
\d(\sqrt{-g}R)=E(g)\d g +\pa_{\mu}\Theta^\mu (\d g, g), 
$$
and  $E(g)=0$ for a solution of the Einstein equations.
Other forms of the presymplectic structures \cite{BB,BC} differ
from the Lee-Wald form by the terms vanishing on solutions of the linearized 
Einstein equations.
A symplectic 2-form is defined as an integral over a codimension-1 
spacelike surface
$$
\Omega^{LW} (\d_1 g,\d_2 g,g)=\int_{\S}w^{\mu\,LW}(\d_1 g,\d_2 g,g)d^3 x_{\mu}.
$$
The Lee-Wald presymplectic form  contracted with a metric  perturbation 
generated by a vector field $\xi^{\mu}$ and any metric variation $\d g$
from the tangent space of metrics satisfies the on-shell relation
$$
 w^{LW} (\d g, \d_{\xi}g,g) \simeq d K_{\xi}^{IW} (\d g, g),  
$$
where $K_{\xi}^{IW}$ is the Iyer-Wald surface charge form 
\cite{IW,WZ} and the equality
is valid up to terms vanishing on-shell.
Variation of the surface charge associated with a transformation
generated by a vector field $\xi^{\mu}$ is defined as
$$ 
\d H_{\xi} =\oint_{\pa \S} K^{IW}_{\xi},   
$$
where integration is over a surface enclosing the horizon 
\cite{LW,IW,BB,BC,WZ,mao}.

The black hole solutions containing supertranslation field 
 are obtained from the Schwarzschild
solution (in isotropic spherical coordinates) by applying the "large"
transformations containing supertranslation field. 
Supertranslation field is a real function on the unit sphere.
The event horizon of a metric containing supertranslation field 
constructed in \cite{comp2}  ($\r$-system) is located at a surface 
which depends on supertranslation field. 
It is possible to construct another coordinate system
("$r$'-system), connected
to the $\r$-system by a "large" transformation, in which
 the horizon is located at
the surface $r=2M$, where $M$ is the mass of black hole.

We calculate  the surface charge forms $K_{\xi}^{\m\n}$ 
in different coordinate systems
connected by  "large" transformations and also within $\r$ and $r$ systems
in coordinate systems corresponding to different parametrizations 
of the unit sphere on which  supertranslation field is defined.  

To obtain the surface charge $H_{\xi}$,  variation $\d H_{\xi}$
should be integrated over the space of metrics.  The unique surface
charge is obtained, if the integral over the space of metrics is
independent of a path of integration.
We find that in the general case variation of the surface charge 
$\d H_{\xi}$ cannot be written as a
variation of a certain  functional over the space of metrics, and 
integration over the space of metrics does not yield a path-independent charge.
We discuss a special case in which  the surface charge of 
horizon symmetries is  obtained in the closed form. 

The paper is organized as follows.

In Sect.2 we review the form of the static vacuum metric 
containing supertranslation field in a $\r$-system obtained 
in \cite{comp2}. Next, 
by a "large" transformation we transform the metric to the $r$-system.
In both $\r$ and $r$-systems we obtain the metrics in different
parametrisations of the unit sphere on which is defined supertranslation
field.

In Sect.3  we study  diffeomorphisms preserving the near-
horizon form of the metric in  $\r $ and $r$-systems. We find
constraints on the generators of transformations 
preserving the gauge and the near-horizon form of the metric.

In Sect.4 we  consider supertranslations preserving the gauge and
near-horizon structure of metric which are extendable in the bulk.
Supertranslations form a group under the modified bracket \cite{comp2}.
A case of supertranslation field depending only on an angle $\th$
 is considered in detail. It is shown that in the case of supertranslation field
depending only on $\th$ the requirement that supertranslation preserves the gauge 
and the form of the metric fixes parameter of transformation through the 
supertranslation field $C(\th )$.   

In Sect.5 we calculate variations of the surface charge corresponding
to  horizon  symmetries  in the $\r$ and $r$-systems.
Variation of the charge is obtained by integration of  
 the  charge surface forms over  the surfaces enclosing the horizon 
and located at a distance $x$ from the horizon.

In the $\r$-system, 
because of the specific form of the surface enclosing the horizon,
 the variation of the charge $\d H_\xi$
 receives  contributions from integrals of the surface charge densities $K_{\xi}^{\m\n}$
 with different components $(\m ,\n) = (t,\r ), (t,\th ), (t,\p)$. 
In the $r$-system the only contribution is from integration of the component with
$(\m\n)=(r,t)$ over the unit sphere.

In the charge densities we separate the leading in $x$ terms
in accordance with the power of $x$ coming from the determinant 
of metric so that the resulting expression for the variation of the charge
is independent on $x$.

In Sect. 6 we discuss integrability of the variation of 
the surface charge in the case of supertranslation field
depending only on a spherical angle $\th$. In the $\r$-system, the 
variation of the charge cannot be presented as a variation of a functional
over the space of metrics. In the $r$-system we
 show that the variation of the charge is
integrable. Although the variations of the charge have different forms 
in the $\r$ and $r$-systems, performing the change of coordinates, we
show that the the expressions are equal.

The last section contains a brief summary of results.

\section{Static vacuum solution of the Einstein equations with 
supertranslation field}
\setcounter{equation}{0}
\renewcommand{\theequation}{2.\arabic{equation}}

We begin this section with a short review of a black hole solution with a
supertranslation field constructed in \cite{comp2}. Next, we transform the metric
to a form in which the horizon is located at the surface $r =2M$.

The vacuum solution of the Einstein equations containing supertranslation field
$C(z^a )$ is
\ba\nonumber
&{}& ds^2 ={g}_{tt}dt^2 +{g}_{\r\r}d\r^2 +{g}_{ab}dz^a dz^b   = \\ \l{2.1}
&{}& -\f{(1-M/2\r_s )^2}{(1+M/2\r_s )^2}dt^2 +
(1+M/2\r_s )^4 \left[d\r^2 + (((\r -E)^2 +U)\g_{ab} +(\r -E)C_{ab})dz^a dz^b
 \right],
\ea
where $z^a$ are coordinates on the unit sphere.
Supertranslation field  $C(z^a )$ is
a real regular function on the unit sphere. Coordinates on the sphere $z^a$ can be
 realized as spherical coordinates $\th ,\p$ with the metric
 $ds^2 =d\th^2 +\sin^2\th d\p^2$, or as projective coordinates
$z^1 =z = \cot\f{\th}{2} e^{i\p}, z^2 =\z =\cot\f{\th}{2} e^{-i\p} $ with the metric
$ds^2 =2\g_{z\b{z}}dz d\b{z},\,\, \g_{z\z}= 2e^{-2\psi},\,\, \psi =
\ln(1+|z|^2 )$.
Covariant derivatives $D_a $ are  defined
with respect to the corresponding metric on the sphere.
Here
\be
\l{2.2}
\r_s (\r, C) =\sqrt{(\r -C -C_{00})^2 + D_a C D^a C}
.\ee
$C_{00}$ is the lowest spherical harmonic mode of $C(z^a )$.
In the following we do not write $C_{00}$ explicitly understanding
$C\rightarrow C-C_{00}$.
The horizon of metric (\ref{2.1}) is located at the surface $\r_s =M/2$.
Here $\r \subset (0, +\infty)$.
The tensor $C_{ab}$  and the functions
$U$ and $E$ are defined as
\ba
\l{2.3}
\nonumber
&{}&C_{ab} = -(2D_a D_b -\g_{ab} D^2 ) C,\\\nonumber
&{}&U=\f{1}{8} C_{ab}C^{ba},\\
&{}&E=\f{1}{2}D^2 C + C,
\ea
The metric (\ref{2.1}) in coordinates $(\r , \th ,\p )$ with supertranslation
field $C(\th ,\p )$
 was obtained from the Schwarzschild metric 
$$
ds^2 =-\left(\f{1 -M/2\r_s}{1+M/2\r_s}\right)^2 dt^2 +\left(1 +M/2\r_s \right)^4
 (dx_s^2 + dy_s^2 +dz_s^2 )
$$
$$
\r_s^2  =x_s^2 + y_s^2 + z_s^2
$$
by the diffeomorphism \cite{comp2}
\ba
\nonumber
&{}&x_s = (\r -C)\sin\th\cos\p + \pa_\p C \sin\p/\sin\th -\pa_\th C \cos\th\cos\p
,\\\nonumber
&{}&y_s	= (\r -C)\sin\th\sin\p - \pa_\p	C \cos\p/\sin\th -\pa_\th C \cos\th\sin\p
,\\\l{m.1}
&{}&z_s = (\r -C)\cos\th -\pa_\th C \sin\th
.\ea
In coordinates $(\r ,\th ,\p )$ the transformed metric is
\ba
\nonumber
&{}&ds^2 =g_{tt}dt^2 +g_{\r\r}[d\r^2 +2\t{g}_{\th\p}d\th d\p +\t{g}_{\th\th}d\th^2 +
\t{g}_{\p\p}d\p^2] =\\\nonumber
&{}& =-\f{(\r_s-M/2)^2}{(\r_s+M/2 )^2}dt^2 +
(1+M/2\r_s )^4 \left[ d\r^2+ 2(\r-E)C_{\th\p}d\th d\p + \right.\\\l{2.11}
&{}&
+\left.    \left(\r-E+\f{1}{2}C_{\th\th}\right)^2 d\th^2 +
\sin^2 \th \left(\r-E-\f{1}{2}C_{\th\th}\right)^2 d\p^2  \right]
\ea
where
\ba
\nonumber
C_{\th\th}=-C''  +C'\cot\th +\f{\ddot{C}}{\sin^2\th}
\qquad C_{\p\p}=-C_{\th\th}\sin^2\th,
\qquad C_{\th\p}=-2(\dot{C}'-\dot{C}\cot\th ).
\ea
Here dot and prime are derivatives over $\p$ and $\th$.

In variables $\r, z^a $ the metric with supertranslation field $C(z,\z )$ 
has a form 
\ba\nonumber
&{}&ds^2 =g_{tt}dt^2 +g_{\r\r}[d\r^2 +\t{g}_{ab}dz^a dz^b ]= \\\nonumber
&{}&= -\f{(1-M/2\r_s )^2}{(1+M/2\r_s )^2}dt^2 +
(1+M/2\r_s )^4 [d\r^2 + 2((\r -E)^2 +U)\g_{z\z}dzd\z +\\\l{2.13}
&{}& (\r -E)(C_{zz} dzdz+C_{\z\z} d\z d\z )]
,\ea
where
$$
C_{zz}=-2D_z\pa_z C,\qquad C_{\z\z}=-2D_{\z} \pa_{\z} C, \qquad C_{z\z}=0.
$$
Transformation from $(\th ,\p ,\r )$ to $(z,\z ,\r )$ is a pure gauge
transformation.

If the supertranslation field depends only on $|z|$, or in coordinates $(r, \th ,\p)$
only on $\th$, the metric simplifies with $g_{\th\p}=0$ (\ref{2.11}). On the other hand, if
the supertranslation field depends only on $z/\z$, or on $\p$, the metric retains its
general form.

Next, we transform the metric (\ref{2.13}) to new variables $(r, z^a)$, where $z^a =(z,\z)$
or $(\th ,\p )$.
The new variables are chosen so that in new variables the  ${tt}$
component of metric is equal to \cite{iofa}
$$
g_{tt}= 1-2M/r\equiv V
.$$
A variable $r\geq 2M$ is defined through the variables $\r ,\th ,\p$  by the relation
\be
\l{2.4}
r=\r_s (\r, C)\left(1+\f{M}{2\r_s (\r, C)}\right)^2
.\ee
Inversely,  $\r$ is expressed through $r$ as
\ba
\l{2.5}
 \r=C +\sqrt{\f{K^2}{4}-D_a C D^a C}= C+\f{K}{2}\sqrt{1 -b_a b^a},
\ea
where we introduced the functions
\be
\l{2.6}
K=r-M + r V^{1/2},\qquad b_a = \f{2D_a C}{K} \qquad b^2 =b_a b^a .
\ee
Differential $d\r (r, z^a)$ is
\ba
&{}&\nonumber
d\r=\r_{,a}dz^a +\r_{,r}dr \\\l{2.7}
&{}&
\r_{,a}= \f{K}{2\sqrt{1-b^2} }
\left(b_a\sqrt{1-b^2 } -\f{\pa_a b^2}{2}\right)  ,
\qquad\r_{,r}=\f{K}{2\sqrt{1-b^2}\, rV^{1/2}}.  
\ea
Using the relations
\be
\l{2.10}
{g}_{tt}=\f{(1-M/2\r_s )^2}{(1+M/2\r_s )^2}=V, \qquad
{g}_{\r\r}=(1+M/2\r_s )^4 =\f{4r^2}{K^2}, \qquad 
\ee
and introducing the transformed metric components ( in variables $(r, z^a)$
the metric components are written with hats).
\ba
\l{2.111}
\h{g}_{rr} = {g}_{\r\r}\r_{,r}^2 =\f{1}{V(1-b^2 )},\qquad
\h{g}_{ra} =  {g}_{\r\r}\r_{,r}\r_{,a},\qquad
\h{g}_{ab}={g}_{\r\r}(\t{g}_{ab}+\r_a\r_b),
\ea
we obtain the metric in a form
\ba
\nonumber
&{}&
ds^2 =\h{g}_{tt}dt^2 +\h{g}_{rr}dr^2 + 2\h{g}_{r a}dr dz^a + \h{g}_{ab}dz^a dz^b =\\
\l{2.9}
&{}&= -Vdt^2 +\f{4r^2}{K^2}[\r_r^2 dr^2 +2\r_r \r_a dr dz^a + 
(\t{g}_{ab}+\r_a \r_b )dz^a dz^b].
\ea

For the above expressions to be well-defined, we require
that $1-b^2 >0$. Because $K$ is an increasing function of $r$
which has its minimum at $r=2M$, the  sufficient condition is
$1-(2DC/M)^2>const>0$.
In the following we work in the units $M=1$.


\section{Diffeomorphisms preserving the near-horizon form of the metric}
\setcounter{equation}{0}
\renewcommand{\theequation}{3.\arabic{equation}}

\subsection{The metric in variables $\r ,z^a$ }

In this section we study  diffeomorphisms which preserve the near-horizon
form  and the gauge of the metric (\ref{2.13}) in the $\r$-system.
Near-horizon foliation of of the space-time is done as follows. 
Let $x$  be a parameter specifying the distance
from a near-horizon to horizon surface (the choice of $x$ depends
on the choice of coordinates and is specified below).
Horizon is located at the surface $\r_s =1/2$ ($\r_s$ is defined in (\ref{2.2})).
Near-horizon surface is defined as a surface $\r_s =1/2 +x$.
Assuming that at the horizon the equation $\sqrt{(\r -C)^2 +(D_a CD^a C}  = 1/2$
has a unique solution, 
\be
\l{3.1}
\r_H (z,\z ) =C+\sqrt{1/4 -D_a C D^a C},
\ee 
by continuity the equation 
$\r_s =\sqrt{(\r -C)^2 +D_a CD^a C}  = 1/2 +x $ 
in some vicinity of $\r_s =1/2$ also has a unique solution 
$$ \r (x,z^a ) = C +\sqrt{(1/2+x)^2 -D_a CD^a C}.$$
 For a small $|x|\ll 1$ we obtain
\be
\l{3.2}
\r (x,z^a )\simeq \r_H (z^a ) +\t{x}
,\ee
where 
\be
\l{3.3}
\t{x}=\f{x}{2\sqrt{1/4 -(DC)^2 }}
.\ee
There are two branches of $\r :\, \r =C\pm\sqrt{\r_s^2 -(DC)^2 }$.
To have a smooth limit to the Schwarzschild metric, $C\rightarrow 0$,
we choose the plus sign. 

In the near-horizon region the metric has a form
\be
\l{3.4}
ds^2 =(\b{g}_{tt}\t{x}^2 + O(\t{x}^3 ))dt^2 +(\b{g}_{\r\r} +
 O(\t{x}))d\r^2 +(\b{g}_{ab}+O(\t{x}))dz^a dz^b 
.\ee
Here $\b{g}_{\m\n}$ are the $O(\t{x}^0 )$ parts of ${g}_{\m\n}$.

We consider transformations generated by the vector field
\be
\l{3.5}
\xi^i =\xi^t\pa_t +\xi^\r\pa_\r + \xi^a\pa_a
.\ee
$\xi^i$ are assumed to be independent of $t$.
General near-horizon transformations are required to
 preserve the gauge of the metric and
the power  $\t{x}^n$ of the leading in $\t{x}$ terms
 in the difference $g_{\m\n}(\t{x})-g_{\m\n}(0)$
 up to a numerical coefficient at the leading term in $\t{x}$.

The metric is written in the gauge ${g}_{\r a}={g}_{\r t}={g}_{ a t}=0$.
Transformations which preserve the gauge satisfy the relations
\ba
\l{3.6}
L_\xi {g}_{\r a}=\pa_{\r} \xi^b {g}_{ba} +\pa_a \xi^{\r}{g}_{\r\r}=0,\\\l{3.7}
L_\xi {g}_{\r t}=\pa_{\r} \xi^t {g}_{tt} +\pa_t \xi^{\r}{g}_{\r\r}=0,\\\l{3.8}
L_\xi {g}_{a t}=\pa_a \xi^{t}{g}_{tt} +\pa_{t} \xi^b {g}_{ba}=0.
\ea
Conditions  (\ref{3.7}) and (\ref{3.8}) give $\xi^t =const$. 

At the near-horizon surface  $\r_s =1/2 +x$, or $\r=\r_H +\x $, and
 the component ${g}_{tt}$ is
\be
\l{3.9}
{g}_{tt} \simeq -4(\r_H -C)^2\x^2 +O(\x^3 )
.\ee
Under the action of transformation generated by vector field $\xi^k$ the component
 ${g}_{tt}$ is transformed as
\be
\l{3.10}
L_\xi {g}_{tt} =-4\f{\r_s -1/2}{(\r_s +1/2)^3}L_\xi\r_s
,\ee
where
\ba
\l{3.11}
L_\xi\r_s =(\xi^\r \pa_\r +\xi^a D_a )\r_s =\f{\xi^\r 2(\r -C) +
\xi^a(-2\left(\r-C) D_a C+ D_a (D_b C D^b C )\right)}{2\r_s}
.\ea
To preserve the near-horizon behavior of $g_{tt}$ (\ref{3.9}), 
 it is necessary that
\be
\l{3.121}
\xi^\r 2(\r -C) +
\xi^a(-2(\r-C) D_a C+ D_a (D_b C D^b C )=O(\t{x})
.\ee
At the horizon, this condition gives the relation connecting $\xi^\r$ and $\xi^a$ 
\be
\l{3.12}
\xi^\r (\r_H -C) +
\xi^a((-\r_H +C) D_a C+ (D_a D_b C) D^b C)|_{\r=\r_H }=0
.\ee
Because $g_{\r\r}$ is a function of $\r_s$, condition (\ref{3.121}) also
ensures that $L_\xi g_{\r\r} = O(\t{x})$.
 
\subsection{The metric in variables $r, z^a$ }

In variables $(r, z^a )$ the horizon of the metric (\ref{2.9}) is located at the surface
$r=2$. 
In the foliation of the near-horizon region through $\r_s =1/2 +x$,
 the near-horizon surface is at $r=2+\h{x}$, where $ \h{x}=2x^2+ O(x^3 )$.
In the near-horizon region  the metric (\ref{2.9}) has a form
\ba
\nonumber
&{}&ds^2 = \h{g}_{tt}dt^2 +\h{g}_{rr} dr^2 +2\h{g}_{ra}dr dz^a +\h{g}_{ab}dz^a dz^b=
\\\nonumber
&{}&=(-\b{g}_{tt} \h{x} +O(\h{x}^2 ))dt^2 + \left(\f{\b{g}_{rr}}{\h{x}} 
+O(\h{x}^{-1/2} )\right)d\h{x}^2+
2\left(\f{\b{g}_{ra}}{\h{x}^{1/2}}+O(\h{x}^0 \right)d\h{x}dz^a + \\\l{3.13}
&{}&+(\b{g}_{ab,0}\h{x}^0+O(\h{x}^{1/2})dz^a dz^b .
\ea
$\b{g}_{\m\n}=O(\h{x}^0 )$ are the coefficients  
at the leading in $\h{x}$ terms in the metric components.
The metric (\ref{3.13}) is written in the gauge 
$$
\h{g}_{rt}=\h{g}_{ta}=0
.$$
The near-horizon transformations are generated by the vector fields
\be
\l{3.14}
\c^k = \c^t\pa_t +\c^r\pa_r +\c^a\pa_a
.\ee
Transformations preserving the gauge conditions are
\ba
\nonumber
L_{\c}\h{g}_{rt}=\pa_r \c^t \h{g}_{tt} +\pa_t\c^r \h{g}_{rr}+\pa_t \c^a \h{g}_{ar}=0\\
\l{3.15}
L_{\c}\h{g}_{at}=\pa_a \c^t \h{g}_{tt}+\pa_t\c^r \h{g}_{ra}+\pa_t \c^b \h{g}_{ba}=0
.\ea
From conditions (\ref{3.15}) we obtain that $\c^t =const$.
Transformations preserving the leading in $\h{x}$ behavior of the metric
components are
\ba
\nonumber
&{}& L_\c \h{g}_{tt} = \c^r\pa_r \h{g}_{tt} =O(\h{x}),\\\nonumber
&{}& L_\c \h{g}_{rr}= \c^r\pa_r \h{g}_{rr}+\c^a\pa_a  \h{g}_{rr} +
2\pa_r \c^a \h{g}_{ar} + 2\pa_r \c^r \h{g}_{rr} =O(\h{x}^{-1}),\\\nonumber
&{}& L_\c \h{g}_{ar}=  \c^r\pa_r \h{g}_{ar}+\c^b\pa_b \h{g}_{ar}
+ \pa_a \c^r \h{g}_{rr} +
\pa_a \c^b \h{g}_{br} + \pa_r \c^r \h{g}_{ar}
 + \pa_r \c^b \h{g}_{ab} =O(\h{x}^{-{1/2}}),\\\l{3.16}
&{}& L_\c \h{g}_{ab } =\c^r\pa_r\h{g}_{ab} +\c^c\pa_c \h{g}_{ab}+
\pa_a \c^r \h{g}_{rb} +
\pa_a \c^c \h{g}_{cb} +\pa_b \c^r \h{g}_{ra} + \pa_b \c^c \h{g}_{ca}=O(\h{x}^0 ).
\ea
We look for the components of the vector field $\c^k$ in a form of expansion
in powers in $\h{x}^{1/2}$
$$
\c^k = \c^k_0 +\h{x}^{1/2}\c^k_{1/2} +\h{x} \c^k_{1} +\cdots .
$$
From the relations (\ref{3.16}) we obtain the form of generators preserving
the near-horizon form of the metric
\be
\l{3.17}
\c_0^r =\c_{1/2}^r =0,\qquad\c^r= \c_1^r \h{x},\qquad
\c^a =\c^a_0 +\h{x}^{1/2}\c^a_0
.\ee
The vector fields generating the near-horizon transformations form the Lie brackets
\be
\l{3.18}
[\c_{(1)}, \,\c_{(2)}]^k = \c_{(12)}^k
,\ee
where
\ba\nonumber
&{}&\c^r_{(12),1}=
\c^b_{(1),0}\stackrel{\leftrightarrow}\pa_b \c^r_{(2),1},
\\\nonumber
&{}& \c^a_{(12),0}=
\c^b_{(1),0}\stackrel{\leftrightarrow}{\pa}_b\c^a_{(2),0},\\\l{3.19}
&{}& \c_{(12),1/2}^a=
\c^b_{(1),0}\stackrel{\leftrightarrow}\pa_b\c^a_{(2),1/2}
+1/2 \left(\c^r_{(1),1}\c^a_{(2),1/2}-
(1\rightarrow 2)\right).
\ea
The vector field (\ref{3.14}) is connected with the vector field
(\ref{3.5}) by a transformation 
\ba
\nonumber
&{}&
\c^r =\xi^\r \f{\pa r}{\pa\r}+\xi^a \f{\pa r}{\pa z^a }=
\xi^\r \f{\pa r}{\pa\r_s}\f{\pa \r_s}{\pa \r } +
\xi^a \f{\pa r}{\pa\r_s}\f{\pa \r_s}{\pa z^a},\\\l{3.20} 
&{}&\c^a =\xi^\r \f{\pa z^a}{\pa\r}+\xi^b \f{\pa z^a}{\pa z^b}=\xi^a.
\ea
From (\ref{2.4}) and (\ref{2.5}), we have
\be
\l{3.21}
\pa r/\pa\r_s =\f{K^2 -1}{K^2},\qquad \pa\r_s/\pa\r =\sqrt{1-b^2},\qquad
\pa\r_s/\pa z^a =\f{K}{4}[-2b_a\sqrt{1-b^2}+D_a b^2 ]
\ee
Using the relations (\ref{3.21}), we obtain
\be
\l{3.22}
\c^t= \xi^t , \qquad \c^r = \f{K^2 -1}{K^2}
\left[\xi^\r \sqrt{1-b^2}+\xi^a \f{K}{4}\left(-2b_a\sqrt{1-b^2}+D_a b^2\right)\right],
\qquad \c^a =\xi^a .
\ee
The expression in the square brackets in $\c^r$ is the same as in (\ref{3.12}). 
For  $|\h{x}|\ll 1$ we have
\be
\l{3.23}
K\simeq 1+\sqrt{2\h{x}},\,\qquad b_a=2\pa_a C (1-\sqrt{2\h{x}}),\qquad
(K^2 -1 )/K^2 =O(\h{x}^{1/2} ).
\ee
At the near-horizon surface the metric component $\h{g}_{tt}$ is
\be
\l{3.24}
\h{g}_{tt}=-\f{\h{x}}{2}+O(\h{x}^2 )
.\ee
To have the transformed metric component $\h{g}_{tt}$  of order $O(\h{x})$, the
vector component $\c^r$ should be of order $O(\h{x})$.
It follows that
\be
\l{3.25}
\xi^\r \sqrt{1-b^2}+\xi^a \f{K}{4}\left(-2b_a\sqrt{1-b^2}
+D_a b^2\right)=O(\h{x}^{1/2}).
\ee
Noting that $\h{x}\sim \t{x}^2$, we see that condition (\ref{3.25})
coincides with the condition (\ref{3.12}).

\section{Supertranslations extended in a bulk: symplectic transformations}
\setcounter{equation}{0}
\renewcommand{\theequation}{4.\arabic{equation}}

In this section we consider supertranslations preserving the near-horizon 
form of the metric which are defined not only in a vicinity
of the horizon, but extend to the bulk. Supertranslations which  preserve
the gauge  of metric (\ref{2.1}) were constructed
in \cite{comp2}.
Supertranslation field in metric (in coordinates $\th ,\p $
transforms under supertranslations as
$$\d_T C(\th ,\p ) =T(\th ,\p ) ,  $$
where $T(\th ,\p )$ is an arbitrary smooth function on the unit sphere.
Generator of supertranslations preserving the static gauge of the 
solution (\ref{2.1})  in coordinates $(\r ,z^a )$  has a form 
\be
\l{4.1}
\xi_T =T_{00}\pa_t -(T-T_{00})\pa_ \r +F^{ab}D_a T D_b
,\ee
where
$$
F^{ab}= \f{C^{ab}-2\g^{ab}(\r -E)}{2((\r -E)^2 -U) }.
$$
Transformations  (\ref{4.1}) are defined in the bulk and  form a commutative
algebra under the modified bracket \cite{comp2}
\be
\l{4.2}
[\xi_1 ,\xi_2 ]_{mod} =[\xi_1 ,\xi_2 ] -\d_{T_1}\xi_2 +\d_{T_2}\xi_1 .
\ee
It is explicitly verified that
\be
\l{4.3}
 \xi^k_T \f{\pa\r_s}{\pa x^k} = \d_T \r_s ,
\ee
where
\be
\l{4.4}
\d_T \r_s (C)= \lim_{\e\rightarrow 0} 
[\r_s (C+\e T) -\r_s (C)]/\e ,
\ee
and
\be
\l{4.5}
\d_T {g}_{\r t}=\d_T {g}_{at}=0
.\ee
General transformations (\ref{4.1}) do not respect the near-horizon form of the metric 
(\ref{3.4}) changing the component $g_{tt}$.
To preserve the near-horizon form of the metric (\ref{3.4}), transformation generated 
by (\ref{4.1}) must satisfy condition (\ref{3.121}). 

If the supertranslation field  depends 
only on $\th, \,\, C=C(\th )$,
the generator of  supertranslations simplifies to
\be
\l{4.6}
\xi_T = T_0\pa_t -(T-T_0 )\pa_\r-\f{T'}{\r-C-C''}\pa_\th
.\ee
The near-horizon structure of the metric is preserved
provided the parameter of transformation (\ref{4.6}) $T(\th )$  satisfies the relation
\be
\l{4.7}
-T(\r_H -C) +T' C' = O(\t{x}) .
\ee
At the horizon, condition (\ref{4.7}) is an ordinary differential equation
on $T(\th )$ with the solution
\be
\l{4.8}
T(\th )=a\exp{\int^\th d\th \sqrt{1/4 -{C'}^2}/C'}
,\ee
where $a$ is an integration constant.
Generators of supertranslations in coordinates $(\r ,x^a )$ and $(r,z^a )$
are connected by the transformation (\ref{3.20}). In variables $(r,z^a )$
generator of supertranslations is
\ba
\nonumber
&{}&
\c_T =\c_T^t\pa_t +\c_T^r\pa_r +\c_T^a \pa_a =\\\l{4.9}
&{}&
 =T_{00}\pa_t + \f{K^2 -1}{K^2}\left(-T\sqrt{1-b^2} +\f{K}{4}F^{ab}D_b T
(-2b_a\sqrt{1-b^2}+D_a b^2)\right)\pa_r +F^{ab}D_b T\pa_a ,
\ea
where in $F^{ab}$ it is substituted $\r -C = K(1-b^2 )^{1/2}/2$
Acting by the generator of supertranslations on the component $\h{g}_{tt}$, we obtain
\be
\l{4.10}
L_{\c_T}\h{g}_{tt} = \f{2}{r^2}\f{K^2-1}{K^2}\left(-T\sqrt{1-b^2} +\f{K}{4}F^{ab}D_b T
(-2b_a\sqrt{1-b^2}+D_a b^2)\right)
.\ee
In the near-horizon region the relations for $K$ are  (\ref{3.23}).
To preserve the form of $\h{g}_{tt}$, it is necessary that
\be
\l{4.11}
-T\sqrt{1-b^2} +\f{K}{4}F^{ab}D_b T
(-2b_a\sqrt{1-b^2}+D_a b^2)=O(\h{x}^{1/2}).
\ee
 This imposes condition on $T(z,\z )$
\be
\l{4.12}
[-T\sqrt{1-b^2} +\f{1}{4}F^{ab}D_b T
(-2b_{a}\sqrt{1-b^2}+D_a b^2)]_{r=0}=0
.\ee
Eq.(\ref{4.12}) for $T$  is solved in Appendix A.
In the case of supertranslated field in  metric depending only on $\th$,
relation (\ref{4.11}) turns into (\ref{4.7}).
It is seen that  in the near-horizon region in variables $r, z^a $ the generator of 
supertranslations has the following structure
\be
\l{4.13}
\c_T =O(x^0 )\pa_t +O(x)\pa_x +O(x^0 ) \pa_a
.\ee

\section{Surface charge of asymptotic horizon symmetries}
\setcounter{equation}{0}
\renewcommand{\theequation}{5.\arabic{equation}}

\subsection{Variables $\r, z^a $.}


In this section,  we calculate the variation of the surface charge
 corresponding to diffeomorphisms preserving the near-horizon
form of the metric. 
Calculations are performed both in $\r$ and $r$-systems.
Variation of the surface charge associated with a symmetry generated 
by a vector field $\xi^\m$ is 
\be
\l{5.1}
\not{\d}H_{\xi} (g,h) =
\f{1}{4\pi}\int_{\pa\S_{\r}} (d^2 x)_{\mu\nu}\sqrt{-g}K_{\xi}^{\mu\nu},
\ee
where $(d^2 x)_{\m\n}=(1/4)\e_{\a\bt\m\n}dx^{\a} dx^{\bt}$. 
 The charge density is
\be
\l{5.2}
K_{\xi}^{\mu\nu}=
\xi^{\mu}\nabla^{\nu} h -\xi^{\mu}\nabla_{\s} h^{\nu \s} +\xi_{\s}\nabla^{\mu} h^{\nu\s}+
\f{1}{2}h\nabla^{\mu}\xi^{\nu} -h^{\mu \s}\nabla_{\s}\c^{\nu}  
+\f{\a}{2}h^{\mu\s}(\nabla^{\nu}\c_{\s}+\nabla_{\s}\c^{\nu})+ (\mu\leftrightarrow \nu) 
,\ee
where $\a =1$ in the Barnich-Brandt form \cite{BB} and $\a =0$ 
in the Iyer-Wald form \cite{IW}.
Here $\pa\S_{\r}$ is a codimension 2 compact spacelike  surface 
$\r -C-\sqrt{(1/2 +x)^2 +(DC)^2}=0 $ enclosing 
the horizon surface. The metric variations  are denoted as
$\d g_{\m\n} \equiv h_{\m\n}$, the inverse  variations are defined as
$h^{\m\n}=g^{\m\r }h_{\r\la }g^{\la\n}$, 
and the trace of metric variations is $h =h_{\m\n}g^{\m\n}.$

First, we consider parametrization of the unit sphere by variables $z, \z$.
Because the metric is static, to the variation of the surface charge 
({\ref{5.1}) contribute integrations over $(z,\z ) ,\,(\r ,z)$ 
and $(\r ,\z )$ 
\be
\l{5.n1}
\int\sqrt{-g}\e_{t\r z\z}dz\w d\z K^{t\r} ,
\qquad\int\sqrt{-g} \e_{t\z \r z}d\r\w dz K^{t\z} ,\qquad
\int\sqrt{-g}\e_{tz \r\z}d\r\w d\z K^{tz}
.\ee
Because at the surface $\S_{\r}$ we have $\r=\r (z^a)$, we rewrite (\ref{5.n1}) as 
\be
\l{5.n2}
\not\d H_\xi =\f{1}{4\pi}\int dz\w d\z\sqrt{-g}\left[K^{t\r}-\r_{,\z}K^{t\z}
-\r_{,z}K^{tz}\right]
.\ee
The charge density $K^{\r t}_\xi$ in the Iyer-Wald form is
\be
\l{5.3}
K_\xi^{\r t}=
\xi^\r\nabla^t h -\xi^\r\nabla_\s h^{t\s} +\xi_\s\nabla^\r h^{t\s}+
\f{1}{2}h\nabla^\r\xi^t -h^{\r\s}\nabla_\s \xi^t 
 -( \r \leftrightarrow t).
\ee
From (\ref{2.13}) we have
$$
g (\r ,z,\z ) =g_{tt}g_{\r\r}^3 \t{g}^{(2)},
$$
where
$$
\t{g}^{(2)} = \t{g}_{zz}\t{g}_{\z\z}-\t{g}_{z\z}^2 = \g_{z\z}^2 [(\r -E)^2 -U]^2
.$$
In variables $\r ,\th ,\p$ expressions (\ref{5.n1})-(\ref{5.3}) have the same 
functional  form as above with the formal change $(z, \,\z)\rightarrow \th ,\p$,
but $\t{g}^{(2)}(\r,\th ,\p )$ is
$$
\t{g}^{(2)} = \t{g}_{\th\th}\t{g}_{\p\p}-\t{g}_{\th\p}^2
, $$
and
$$
\not\d H_\xi =\f{1}{4\pi}\int d\th\w d\p\sqrt{-g}\left[K^{t\r}-\r_{,\p}K^{t\p}
-\r_{,\th}K^{t\th}\right]
.$$
At the surface $\r =\r_H (z^a )+\t{x}$ we have $\sqrt{-g}=O({\t{x}})$, 
and to obtain a non-zero result
for $\not\d H$, in  $K^{\r t}_\xi$ we look for the terms of order $O({\t{x}}^{-1})$.

The five contributions in  $K_{\xi}^{\r t}$ are 
\ba
\nonumber
&{}& 1.\quad\xi^\r\nabla^t h-\xi^t\nabla^\r h =\xi^{\r} g^{tt}\pa_t h -
\xi^t g^{\r\r}\pa_\r h , \\\nonumber
&{}& 2.\quad -\xi^\r\nabla_s h^{ts} +
\xi^t \nabla_s h^{\r s},
       \\\nonumber
&{}& 3.\quad \xi_s \nabla^\r h^{ts}-\xi_s \nabla^t h^{\r s},
\\\nonumber
&{}& 4.\quad  \f{h}{2}(\nabla^\r\xi^t -\nabla^t\xi^\r )=
\f{h}{2}(g^{\r s}\nabla_s \xi^t -g^{ts}\nabla_s\xi^\r ),
\\\l{5.5}
&{}& 5.\quad - h^{\r s}\nabla_{s}\xi^t +  h^{ts}\nabla_{s} \xi^{\r} 
.\ea

Because $h$ is independent of $t$ and $g^{\r\r} =O(\x^0 )$,
 the two terms in the item 1 are of order $\t{x}^0$.

The first term in the item 2 
\ba\nonumber
&{}&-\xi^\r \nabla_s h^{ts}=
-\xi^\r(\nabla_t  h^{tt}+\nabla_{\r}  h^{t\r}+\nabla_a  h^{ta})
=-\xi^\r (2\G^t_{tt} h^{tt}+\G^t_{\r\r} h^{\r\r}+\G^{\r}_{t\r} h^{tt}+
\G^t_{aa} h^{aa}+\G^a_{at} h^{tt})=0
\ea
is zero, because all $\G$  vanish.  
The second term in the item 2 is
\be\nonumber
\xi^t\nabla_s h^{\r s}=\xi^t (\nabla_t  h^{\r t}+\nabla_{\r}  h^{\r\r}+
\nabla_a  h^{\r a})=\xi^t (\G^\r_{tt}h^{tt}+\G^t_{t\r}h^{\r\r} ) + O(\t{x}^0 )
.\ee
The first term in the item 3 is transformed as
\ba\nonumber
&{}&\xi_s \nabla^{\r} h^{ts}=g^{\r\r}(\xi_t \nabla_\r  h^{tt}+
\xi_\r\nabla_{\r}  h^{t\r}+\xi_a\nabla_{\r}  h^{ta})
+\xi_s g^{\r a}\nabla_{a}h^{ts}
=\\\nonumber
&{}&=g^{\r\r}[\xi_t(\pa_\r h^{tt}+2\G^t_{\r t}h^{tt} ) +\xi_\r (\G^t_{\r \r}h^{\r\r}+
\G^{\r}_{\r t}h^{tt}) +O(x^0 )]=\xi_t g^{\r\r}(\pa_\r h^{tt} +g^{tt}g_{tt,\r}h^{tt})
 +O(\t{x}^0 )
.\ea
The leading in $\t{x}$ terms in this expression cancel
$$
\xi_t g^{\r\r}(\pa_\r h^{tt} +g^{tt}g_{tt,\r}h^{tt})
=  \xi_t g^{\r\r}\left(
-2\f{ \b{h}^{tt} }{\t{x}^3 }+ \f{ \b{g}^{tt} }
{\t{x}^2 }2\t{x}\b{g}_{tt}\f{ \b{h}^{tt} }{\t{x}^2 }
\right)=0,
$$
and the remaining expression is of order $\t{x}^0$.
The second term in the item 3
\ba\nonumber
-\xi_s g^{tt} \nabla_t h^{\r s}=-g^{tt}(\xi_t \nabla_t  h^{\r t}+
\xi_\r\nabla_{t}  h^{\r\r}+\xi_a\nabla_t  h^{\r a})=
-\xi^t(\G^{\r}_{tt}h^{tt} + \G^t_{t\r}h^{\r\r} )+O(\t{x}^0 )
\ea
 cancels the corresponding expression in item 2.

Collecting the items 4 and 5, we obtain
\be
\l{5.6}
{K}^{\r t}_\xi=\nabla_{s}\xi^t\left(\f{{h}}{2}{g}^{\r s}-{h}^{\r s}\right)-
\nabla_s\xi^\r\left(\f{{h}}{2}{g}^{ts}-{h}^{ts}\right)+O(\t{x}^0 )
.\ee
Taking into account the form of the metric in the $\r$-system, we write
${K}^{\r t}_\xi$ as
\ba
\l{5.61}
{K}^{\r t}_\xi=\nabla_{\r}\xi^t\left(\f{{h}}{2}{g}^{\r \r}-{h}^{\r \r}\right)-
\nabla_t\xi^\r\left(\f{{h}}{2}{g}^{tt}-{h}^{tt}\right)+O(\t{x}^0 )
.\ea
The leading in $\t{x}$ part of ${K}^{\r t}$ of order $\t{x}^{-1}$ is
\ba
\l{5.7}
\nonumber
&{}&{K}_{\xi}^{\r t}=\Gamma^t_{\r t}\xi^t\left(\f{{h}}{2}{g}^{\r\r}-{h}^{\r\r}\right)-
\Gamma^\r_{tt}\xi^t\left(\f{{h}}{2}{g}^{tt}-{h}^{tt}\right)=
\\
&{}&=\f{\xi^t}{2}{g}_{tt,\r}({h}{g}^{tt}{g}^{\r\r}-
{g}^{tt}{h}^{\r\r}-{g}^{\r\r}{h}^{tt})
=\f{\xi^t}{2}{g}_{tt,\r}{g}^{tt}{g}^{\r\r}{h}_{ab}{g}^{ab} .
\ea
Here it was used that $\xi^t =const$.
The expression ${h}_{ab}{g}^{ab}$ can be written in a form
\be
\l{5.8}
h_{ab}{g}^{ab}=({h}_{zz}{g}^{zz}+{h}_{\z\z}{g}^{\z\z}+
2{h}_{z\z}{g}^{z\z})=\f{1}{ {g}^{(2)} }({h}_{zz}{g}_{\z\z}
+{h}_{\z\z}{g}_{zz}-2{h}_{z\z}{g}_{z\z})=\f{ \d{g}^{(2)} }{g^{(2)} }
.\ee
Calculating the charge density 
\be
\l{5.31}
K_\xi^{z t}=
\xi^z\nabla^t h -\xi^z\nabla_\s h^{t\s} +\xi_\s\nabla^z h^{t\s}+
\f{1}{2}h\nabla^z\xi^t -h^{z\s}\nabla_\s \xi^t
 -( z \leftrightarrow t)
,\ee
 we note that contribution 
from the item 1 is $O(\t{x}^0 )$, contributions from the items 2 and 3
cancel up to terms  $O(\t{x}^0 )$, and the items 4 and 5 yield
\be
\l{5.n3}
K^{zt}=\f{\xi^t}{2}g_{tt,a}g^{tt}\left( \f{h}{2}g^{za} -h^{za}\right)-
\f{\xi^t}{2}g_{tt,z}g^{zz}\left( \f{h}{2}g^{tt} -h^{tt}\right)
.\ee
Expression (\ref{5.n3}) is transformed to a form
\be
\l{5.n4}
K^{zt}=\f{\xi^t}{2}g^{tt}\left[g_{tt,z}\left( h_{\r\r }g^{\r\r}g^{zz}+
h_{\z\z}{g^{(2)}}^{-1} \right)+
g_{tt,\z}\left(\f{h}{2} g^{z\z }-
h^{z\z}\right)\right]+O(\t{x}^0 )
.\ee
In the same way for the charge density $K^{\z t}$  we have
\be
\l{5.n5}
K^{\z t}=\f{\xi^t}{2}g^{tt}\left[g_{tt,\z}\left( h_{\r\r }g^{\r\r}g^{zz}+
h_{\z\z}{g^{(2)}}^{-1} \right)+
g_{tt,z}\left(\f{h}{2} g^{z\z }-
h^{z\z}\right)\right]+O(\t{x}^0 )
.\ee

In variables $t,\r ,\th ,\p$ we obtain the expressions of the form 
(\ref{5.n4})-(\ref{5.n5}) with $\th ,\p$ substituted for $z, \z$.
Although the metric components have similar functional form, 
the actual expressions written in variables are very different.

Let us consider the case of supertranslation field $C(\th )$ 
depending only on $\th$.
The horizon surface is
$$ \sqrt{(\r_H -C)^2 +{C'}^2}-1/2 =0
,$$
where  $\r_H =\r (\th )$.
Because $\r_{,\p}=0$, we have
\be
\l{5.n7}
\not\d H_\xi =\f{1}{4\pi}\int d\th\w\d\p\sqrt{-g} 
\left[K^{\r t}-\r_{,\th}K^{\th t}\right]
.\ee
The charge density $K^{\r t}$ is
\be
\l{5.n8}
{K}_\xi^{\r t}=
\f{\xi^t}{2}{g}_{tt,\r}{g}^{tt}g^{\r\r}\f{ h_{\th\th}{g}_{\p\p}+h_{\p\p}
{g}_{\th\th}}{{g}^{(2)} },
\ee
where ${g}^{(2)}={g}_{\th\th}{g}_{\p\p}$ .
For the density $K^{\th t}$ we obtain
\be
\l{5.n9}
K^{\th t} =\f{\xi^t}{2}g^{tt}g_{tt,\th}\left[ h_{\r\r}g^{\r\r}g^{\th\th}+
h_{\th\th}g^{\th\th}g^{\p\p} \right]
.\ee
The term with $h_{\r\r}$ vanishes and does not contribute to $\d H_{\xi}$.
Variation of the surface charge is
\be
\l{5.n11}
\not\d H_{\xi } =\lim\limits_{\r_s\to 1/2}\f{1}{4\pi }\int d\th \w d\p
\sqrt{-g}
\f{\xi^t}{2}g^{tt}\left[g_{tt,\r}g^{\r\r}\left(\f{\d g_{\th\th}}{g_{\th\th}}
+\f{\d g_{\p\p}}{g_{\p\p}}\right)-
g_{tt,\th}g^{\th\th}\f{\d g_{\p\p}}{g_{\p\p}}\r_{,\th}\right]
.\ee
Here $g^{\th\th}= \t{g}^{\th\th}g^{\r\r}$. In the near-horizon region $\r_s= 1/2 +x,\,\,
|x|\ll1$ we have
\be
\l{5.n12}
\f{g_{tt,\r}}{\sqrt{g_{tt}}}=2\f{(\r_s-1/2 )(\r-C)}{(\r_s +1/2)} \simeq 4(\r-C),\quad
\f{g_{tt,\th}}{\sqrt{g_{tt}}} \simeq 4(-C')(\r-C-C''),\quad
\f{\r_{,\th}}{\sqrt{\t{g}_{\th\th}}}\simeq \f{C'}{ (\r -C )}
.\ee
We obtain variation of the charge as
\be
\l{5.n13}
\not\d H_{\xi } =\lim\limits_{\r_s\to 1/2}\f{1}{4\pi }\int d\th \w d\p
2\xi^t\left[ (\r-C)\d \sqrt{g_{\th\th} g_{\p\p}}+
 \f{C'}{ (\r -C )} \d\sqrt{ g_{\p\p}}\right]
,\ee
where at the horizon $\r_H -C =\sqrt{1/4 - {C'}^2 }$.


Integrability of variation of the charge means that an integral 
of a variation over the manifold of metrics  is path-independent.
The expression for $\not\d H_{\xi }$, (\ref{5.n13} ) is not of the form 
of a variation of a function over the space of metrics and cannot be 
integrated over the space of metrics in a path-independent way.

\subsection{Variables $r, z^a $.}


Let us turn to calculation of the variation of the surface charge
 in variables $(r,\,z^a)$ with the metric (\ref{2.9})
$$
ds^2 =\h{g}_{tt}dt^2 +\h{g}_{rr}dr^2 + 2\h{g}_{r a}dr dz^a + \h{g}_{ab}dz^a dz^b
.$$
At the surface $r=2+\h{x}$ enclosing the horizon surface $r=2$, the near-horizon forms 
of the metric (\ref{2.9}) and its inverse are
\ba
\l{5.12}
\begin{array}{c} {}\\ \h{g}_{mn}=\\{}\\{}\end{array}
\left|\begin{array}{cccc}\b{g}_{tt} \h{x} &0&0&0\\
0 & \b{g}_{rr}/\h{x} & \b{g}_{rz}/\sqrt{\h{x}} & \b{g}_{r\z}/\sqrt{\h{x}}\\
0 & \b{g}_{rz}/\sqrt{\h{x}} & \b{g}_{zz} & \b{g}_{z\z}\\
0 & \b{g}_{r\z}/\sqrt{\h{x}} & \b{g}_{z\z} & \b{g}_{\z\z}
\end{array}\right|;
\qquad
\begin{array}{c} {}\\ \h{g}^{mn}=\\{}\\{}\end{array}
\left|\begin{array}{cccc}\b{g}^{tt}/ \h{x} &0&0&0\\
0 & \b{g}^{rr}\h{x} & \b{g}^{rz}\sqrt{\h{x}} & \b{g}^{r\z}\sqrt{\h{x}}\\
0 & \b{g}^{rz}\sqrt{\h{x}} & \b{g}^{zz} & \b{g}^{z\z}\\
0 & \b{g}^{r\z}\sqrt{\h{x}} & \b{g}^{z\z} & \b{g}^{\z\z}
\end{array}\right|
,\ea
where $\b{g}_{mn}$ denotes the factor of order $O(\h{x}^0 )$ .
Variation of the surface charge  is
\be
\l{5.13}
\not\d H_\c (\h{g},h) = 
\f{1}{4\pi}\int_{\S_r}(d^2 x)_{rt}\sqrt{-\h{g}} \h{K}^{rt}_{\c}(\d\h{g},\h{g})
\ee
where
\be
\l{5.14}
\h{K}^{IW\,rt}_\c = \c^r\h{\nabla}^t \h{h} -\c^r\h{\nabla}_s \h{h}^{ts} 
+\c_s\h{\nabla}^r \h{h}^{ts}+
\f{\h{h}}{2}\h{\nabla}^r\c^t -\h{h}^{rs}\h{\nabla}_s \c^t- (r\rightarrow t)
\ee
and ${\S_r}$ is a surface $r=2+\h{x}$.
Here $\h{g}= g_{tt}\h{g}^{(3)}$, and $\h{g}^{(3)}$  is determinant of the 
3D part of the metric
\be
\l{5.15}
\h{g}^{(3)}(r,z,\z )=\h{g}_{rr}(\h{g}_{zz}\h{g}_{\z\z }-\h{g}^2_{z\z })
-\h{g}^2_{r z }\h{g}_{\z\z } -\h{g}^2_{r\z }
\h{g}_{zz} + 2\h{g}_{rz }\h{g}_{r\z }\h{g}_{z\z }
.\ee
Using the expressions (\ref{2.5})-(\ref{2.10}), 
 determinant $\h{g}^{(3)}$ can be writen as
\be
\l{5.22}
\h{g}^{(3)}= g_{\r\r}\r_r^2 g^{(2)}=\f{4r^2}{K^2}
\left(\f{K}{2\sqrt{1-b^2} }\f{1}{ rV^{1/2}}\right)^2 (\t{g}_{zz}\t{g}_{\z\z}-
\t{g}^2_{z\z})
\ee
with $\t{g}_{ab}$ from (\ref{2.13}).
Determinant of the metric $\h{g}$  is of order $O(\h{x}^0)$, and to extract a
contribution nonzero at the horizon we must select in $\h{K}^{rt} $ the 
terms of order $O(\h{x}^0 )$.

The five terms in  (\ref{5.14}) are
\ba
\nonumber
&{}& 1.\quad\c^r\h{\nabla}^t \h{h}-\c^t\h{\nabla}^r \h{h} =\c^r \h{h}{g}^{tt}\pa_t \h{h} -
\c^t \h{g}^{rr}\pa_r \h{h} - \c^t \h{g}^{ra}\pa_a \h{h} =O(\h{x}^{1/2}). \\\nonumber
&{}& 2.\quad -\c^r\h{\nabla}_s \h{h}^{ts} +
\c^t \h{\nabla}_s \h{h}^{rs}.
\\\nonumber
&{}& 3.\quad \c_s \h{\nabla}^r \h{h}^{ts}-\c_s \h{\nabla}^t \h{h}^{rs} =
 \c_t \h{\nabla}^r \h{h}^{tt} -\c_t\h{\nabla}^t\h{h}^{rt}-\c_r\h{\nabla}^t  \h{h}^{rr}-
\c_a\h{\nabla}^t  \h{h}^{ra}. 
\\\nonumber
&{}& 4.\quad \f{\h{h}}{2}(\h{\nabla}^r\c^t -\h{\nabla}^t\c^r )=
\f{\h{h}}{2}\left[\h{g}^{rr}\h{\nabla}_r\c^t -\h{g}^{tt} \h{\nabla}_t \c^t\right]
 + O(\h{x}^{1/2}).\\\l{6.16}
&{}& 5. \quad -\h{h}^{rs}\h{\nabla}_s \c^t +\h{h}^{ts}\h{\nabla}_s \c^r =
-\f{1}{2}(\h{h}^{rr}\h{\nabla}_r \c^t   -\h{h}^{tt}\h{\nabla}_t \c^t ) +O(\h{x}^{1/2}).
\ea
Estimating two terms in the item 1, we have
$$
\c^r\h{g}^{tt}\pa_t \h{h}-\c^t( \h{g}^{rr}\pa_r \h{h} + \h{g}^{rr}\pa_a\h{h})=O(\h{x}^{1/2})
$$ 
and the item 1 does not contribute to $\h{K}^{rt}$.

Because all the terms containing  $\G$  one index $t$  are zero,
in the item 2 the first term vanishes
$$
-\c^r  \h{\nabla}_s \h{h}^{ts} =-\c^r (\h{\nabla}_t \h{h}^{tt}+\h{\nabla}_r \h{h}^{tr}+
\h{\nabla}_a \h{h}^{ta})=0.
$$
In the second term 
$$\c^t  \h{\nabla}_r \h{h}^{rr} =\c^t (\h{\nabla}_t \h{h}^{rt}+
\h{\nabla}_r \h{h}^{rr}+\h{\nabla}_a \h{h}^{ra})
$$ 
 the part $\c^t (\h{\nabla}_r \h{h}^{rr}+\h{\nabla}_a \h{h}^{ra})$ is estimated as
\ba
\nonumber
\c^t  \h{\nabla}_r \h{h}^{rr}=\c^t (\pa_r  \h{h}^{rr} +
2\Gamma^r_{rr}\h{h}^{rr} +2\Gamma^r_{ra}\h{h}^{ra})=
\c^t (\b{h}^{rr}+2\Gamma^r_{rr}\h{h}^{rr} +O(\h{x}^{1/2}) )=\\\l{5.17}
=\c^t \left[\b{h}^{rr}+\left(\b{g}^{rr}\h{x}\b{g}_{rr}\left(-\f{1}{\h{x}^2}\right)
+2\b{g}^{ra}\h{x}^{1/2}\b{g}_{ra}\left(-\f{1}{2\h{x}^{3/2}}\right)\right)
\b{h}^{rr}\h{x} +O(\h{x}^{1/2})\right]
=O(\h{x}^{1/2})
.\ea
Because of identity $\b{g}^{rr}\b{g}_{rr}+\b{g}^{ra}\b{g}_{ar}=1$ 
the sum of the terms in round brackets in (\ref{5.17}) is equal to $-\b{h}^{rr}$.
The term $\c^t  \h{\nabla}_a \h{h}^{ra}$ is of order $O(\h{x}^{1/2})$.  In the item 2
there remains the term $\c^t\h{\nabla}_t \h{h}^{rt}$.

In the item 3 the first term is
$$
\c_s\h{\nabla}^r \h{h}^{ts}=\c_t\h{\nabla}^r \h{h}^{tt}+\c_r\h{\nabla}^r \h{h}^{tr}+
\c_a\h{\nabla}^r \h{h}^{ta}=\\
\c_t (\h{g}^{rr}\h{\nabla}_r +\h{g}^{ra}\h{\nabla}_a )\h{h}^{tt} +
\c_r (\h{g}^{rr}\h{\nabla}_r +\h{g}^{ra}\h{\nabla}_a )\h{h}^{tr}+O(\h{x}^0).
$$
The term $\c^t \h{g}_{tt}\h{g}^{ra}\h{\nabla}_a \h{h}^{tt}$  is of order $\h{x}^{1/2}$.
In the term
$$\c^t \h{g}_{tt} \h{g}^{rr}\h{\nabla}_r \h{h}^{tt}=\c^t \b{g}_{tt}\h{x} \b{g}^{rr}\h{x}\left
(-\f{\b{h}_{tt}}{\h{x}^2}+\f{\b{g}^{tt}}{\h{x}}\b{g}_{tt}\f{\b{h}^{tt}}{\h{x}}+O(\h{x}^0 )\right)
$$
the leading-order parts cancel and it 
is also of order $O(\h{x}^{1/2} )$.
  The remaining term in the item 3, equal to $-\c^t\h{\nabla}_t \h{h}^{rt }$, 
cancels the corresponding term in the item 2.

We obtain $\h{K}^{rt}$ as
\ba
\nonumber
&{}& \h{K}^{rt}_\c=\h{\nabla}_s\c^t\left(\f{\h{h}}{2}\h{g}^{rs} -\h{h}^{rs}\right) -
\h{\nabla}_s\c^r\left(\f{\h{h}}{2}\h{g}^{ts} -\h{h}^{ts}\right)+O(\h{x}^{1/2})=
\\\l{5.18}
&{}&=
\f{\c^t}{2}\left[\Gamma^t_{rt}\left(\f{\h{h}}{2}\h{g}^{rr} -\h{h}^{rr}\right)-
\Gamma^r_{tt}\left(\f{\h{h}}{2}\h{g}^{tt} -\h{h}^{tt}\right)
+O(\h{x}^{1/2})\right]
.\ea
Because $g_{tt} =V(r)$, we have $\G^t_{\th t}=\G^{\th}_{tt}=0$.

In $\h{K}^{rt}$ the leading terms are
\be
\l{5.19}
\h{K}^{rt}_\c=
\f{\c^t}{2}\h{g}_{tt,r}[\h{h}\h{g}^{rr}\h{g}^{tt} -{\h{h}}^{rr}\h{g}^{tt}
-\h{g}^{rr}{\h{h}}^{tt}]
 =\f{\c^t}{2}\h{g}_{tt,r}\h{g}^{tt}{\h{h}}_{ab}
(\h{g}^{rr}\h{g}^{ab}-\h{g}^{ra}\h{g}^{rb}).
\ee
Combination in the rhs of (\ref{5.19}) is presented as
\be
\l{5.23}
\h{h}_{ab}(\h{g}^{rr}\h{g}^{ab}-\h{g}^{ra}\h{g}^{rb})=
\f{ \h{h}_{zz}\h{g}_{\z\z}+\h{h}_{\z\z}\h{g}_{zz}-2\h{h}_{z\z}\h{g}_{z\z} }
{\h{g}^{(3)}}
=\f{\d  \h{g}^{(2)} }{ \h{g}^{(3)} },
\ee
where
$$
\h{g}^{(2)}=\h{g}_{zz}\h{g}_{\z\z}-\h{g}_{z\z}^2 .
$$
Variation of the surface charge is
\be
\l{5.24}
\not\d H_{\c} (\h{g},\h{h})=\lim\limits_{r\rightarrow 2}
\f{1}{4\pi}\int d z\w d\z  \sqrt{V\h{g}^{(3)} }
\f{\c^t}{2}\h{g}_{tt,r}\h{g}^{tt} 
\f{\d\h{g}^{(2)} }{\h{g}^{(3)} }.
\ee
Substituting (\ref{5.22}) and taking the limit $r\rightarrow 2$,
we have
\be
\l{5.25}
\not\d H_\c (\h{g},\h{h})=\f{1}{4\pi}\int
dz\w d\z \g_{z\z} \f{\c^t}{2}
\f{ \d\t{\h{g}}^{(2)} }{ \sqrt{\t{g}^{(2)} } } (1/4-D_a CD^a C)^{1/2}
.\ee
"Tilda" indicates that from the expression were extracted povers of $\g_{z\z}$.

The integral (\ref{5.25}) is not of the form of
 a variation of a functional over the space of metrics.
In a general case, expression (\ref{5.25}) is not integrable.
A special case of integrable variation of the surface charge is discussed in
the next section.

\section{Integrable  variations of surface charges}
\setcounter{equation}{0}
\renewcommand{\theequation}{6.\arabic{equation}}

In this section we consider an example of integrable variation of the charge.
We consider  the case of supertranslation field $C(z,\z)$
in coordinate system $r,z,\z$ depending only on $|z|$, or 
in coordinates $r, \th ,\p$, only on $\th$.

Integrability of the charge over the space of metrics means that the charge
$H_\c (g)=\int^g_{\b{g}} \d H_\c $ is independent 
of a form of a path in a space of metrics. 

In coordinates $r,\th ,\p$ the metric (\ref{2.9}) takes a form
\ba
&{}&ds^2 =
\nonumber
-V dt^2 +  \f{dr^2}{V(1-b^2 )} +2drd\th\f{br (\sqrt{1-b^2}-b' )}{(1-b^2 )V^{1/2}}
+\\\l{6.1}
&{}& + d\th^2 r^2\f{ (\sqrt{1-b^2}-b' )^2}{(1-b^2 )} +d\p^2 r^2\sin^2\th (b\cot\th
-\sqrt{1-b^2 })^2
,\ea
where $b= 2C'(\th )/K$.
The charge density $\h{K}^{rt}_{\c}(\d \h{g},\h{g})$  (\ref{5.19}) is
\be
\l{6.2}
\h{K}_{\c}^{rt} =
\f{\c^t}{2}\h{g}_{tt,r}\h{g}^{tt}[\h{h}_{\th\th }
(\h{g}^{rr}\h{g}^{\th\th }-(\h{g}^{r\th })^2 )
+\h{h}_{\p\p }\h{g}^{rr}\h{g}^{\p\p }]
.\ee
Using the relations
\be
\l{6.22}
\h{g}^{rr}\h{g}^{r\th }-(\h{g}^{r\th})^2 =1/{\h{g}}^{(2)}
,\ee
where $\h{g}^{(2)}= \h{g}_{rr}\h{g}_{\th\th }-\h{g}_{r\th }^2$, 
and $\h{g}^{(2)}=\h{g}_{\th\th }/V$ and noting that ${g}^{rr}=V$, we
write $\h{K}^{rt}$ as
\be
\l{6.3}
\h{K}^{rt}_\c =\f{\c^t}{2}\h{g}_{tt,r}\h{g}^{tt}
\left(V\f{\h{h}_{\th\th}}{\h{g}_{\th\th }}+V\f{\h{h}_{\p\p}}{\h{g}_{\p\p}}\right)=
\f{\c^t}{2}V_{,r} \left(\f{\d \h{g}_{\th\th}}{\h{g}_{\th\th}}
+\f{\d \h{g}_{\p\p}}{\h{g}_{\p\p}}\right)
.\ee
We obtain $\not\d H_{\c}$ in a form
\ba
\nonumber
\l{6.4}
\not\d H_\c (\h{g},\d\h{g}) =\lim\limits_{r\rightarrow 2}
\f{1}{4\pi}\int d\th\w d\p
\sqrt{-\h{g}_{tt}\h{g}^{(2)}\h{g}_{\p\p}}\h{K}^{rt}
=\f{1}{4\pi}\int d\th\w d\p\sqrt{\h{g}_{\th\th}\h{g}_{\p\p}}
\f{\c^t}{4}\f{\d (\h{g}_{\th\th}\h{g}_{\p\p})}
{\h{g}_{\th\th}\h{g}_{\p\p}}
.\ea
With $\c^t =const$ (\ref{6.4}) can be written as a variation
\be
\l{6.5}
\not\d H_\c (\h{g},\h{h}) =\f{1}{4\pi}\int_{\S}d\th\w d\p
\f{\c^t}{2}\d\sqrt{\h{g}_{\th\th}\h{g}_{\p\p}}
,\ee
and is integrable. 

Let us consider calculation of $\not\d H_\c (\h{g},\h{h})$ in parametrization
of the sphere in coordinates $(z,\z)$.
In the case $C=C(\th )$ from the relation $\sqrt{\h{g}^{(3)}(r,z,\z )}dz\w d\z
=\sqrt{\h{g}^{(3)}(r,\th,\p )}d\th\w d\p$ it follows that
$$
{\h{g}^{(3)}(r,z,\z )}=\h{g}^{(3)}(r,\th,\p )(\th_{,z}\p_{,\z}-\th_{,\z}\p_{,z})^2
.$$
Expressing the variation $\d (\h{g}_{zz}\h{g}_{\z\z}-\h{g}^2_{z\z})$ through coordinates
$\th ,\p$, we have
$$
\d (\h{g}_{zz}\h{g}_{\z\z}-\h{g}^2_{z\z})=\d (\h{g}_{\th\th}\h{g}_{\p\p})
(\th_{,z}\p_{,\z}-\th_{,\z}\p_{,z})^2
.$$
Substituting $\h{g}^{(3)}(r,\th,\p )=(\h{g}_{rr}\h{g}_{\th\th}-\h{g}_{r\th}^2 )
\h{g}_{\p\p}=\h{g}_{\th\th} \h{g}_{\p\p}/V$, we obtain
the variation of the surface charge in the form (\ref{6.5}).

Let us show that the expression for the variation of the charge
${\not\d} H_\c$
in variables $r,\th ,\p$ is equal
the variation $\not\d H_\xi$ in variables $\r ,\th ,\p $.
Note that the expressions have different functional form, and $\not\d H_\xi$
is not integrable. 
The charge densities in the $r$ and $\r$ systems are connected as
\be
\l{6.m7}
\h{K}^{rt}=K^{\r t}\f{\pa r}{\pa \r}+K^{\th t}\f{\pa r}{\pa\th}.
\ee
The integration measures satisfy the equality
$\sqrt{-\h{g}}dr \w d\th\w d\p=\sqrt{-g}d\r\w d\th\w d\p$ and 
$\sqrt{-\h{g}}dr =\sqrt{-\h{g}}(\pa r/\pa\r) d\r$.
Using the relations
\be
\l{6.m8}
\f{\pa r}{\pa\r}\f{\pa\r}{\pa r}=1, \qquad \f{\pa r}{\pa\r}\f{\pa\r}{\pa \th}+
\f{\pa r}{\pa\th}=0
\ee
and noting that $\c^{t}=\xi^t$, we obtain
\be
\l{6.m9}
\not\d H_\c =\int d\th\w d\p \sqrt{-\h{g}} \h{K}^{rt}=
\int d\th\w d\p\sqrt{-g}
\left[K^{\r t}-\r_{,\th}K^{\th t}\right]=\not\d H_\xi
.\ee
Details of the derivation of the relation (\ref{6.m9}) are contained in
Appendix B.


\section{Summary and conclusions}

In this paper we studied the near-horizon symmetries
of the metric of  black hole containing
  supertranslation field which preserve the near-horizon structure of the
metric. The horizon symmetries were considered in different coordinate systems 
($\r $ and $r$-systems) connected by a "large" diffeomorphism 
containing supertranslation field
 and also in coordinate systems connected
 by a pure coordinate transformation which do not change  supertranslation
field in the metric. 
Foliation of the near-horizon region was defined through a smooth
deformation of the horizon surface $\r_s =1/2 $ to $\r_s =1/2+x$ where 
$\r_s =((\r-C)^2 +(DC)^2 )^{1/2}$.
 In the $r$-system the horizon is is located
at the surface $r=2$ (in units $M=1$), 
in the $\r$-system - at the
surface $\r_H =C+(1/4 +(DC)^2 )^{1/2}$.
In both $\r$ and $r$-systems we constructed diffeomorphisms which preserve the gauge
and the near-horizon form of the metric in the leading order in $x$.

We discussed symplectic transformations which are extendable from the near-horizon region
in the bulk. Symplectic transformations are generated by vector fields which depend
on the supertranslation field in the metric $C(\th ,\p )$ and a field $T(\th ,\p )$ and act
on $C(\th ,\p )$ as $\d_T C(\th ,\p ) = T(\th ,\p )$. In a case of supertranslation field
depending only on $\th$ a condition that transformation preserves the near-horizon form of
metric at the horizon is an ordinary differential equation with a solution for $T$
expressed through $C(\th )$.

We calculated variation of the surface charge in the $\r$ and $r$-systems
and also in different parametrizations of the the unit sphere $\th ,\p$ and $z, \z$. 
In the $r$-system, variation of the charge is expressed through an integral
of the charge density $\h{K}_\c^{rt}(\d\h{g}_{mn}, \h{g}_{mn})$ over the 
sphere. In the $\r$-system, variation of the charge is obtained as a sum
of three integrals over the horizon surface with  charge densities 
$K_{\xi}^{\r t}, K_{\xi}^{\th t}$ and $K_{\xi}^{\p t}$ and corresponding integrations
over $d\th d\p , d\r d\p$ and $d\r d\th$. 
Here $\xi$ and
$\c$ are the vector fields generating asymptotic horizon transformations
 in the $\r$ and $r$-systems.
Surface charge densities were calculated in the leading order in $x$, 
and together with contributions from $\sqrt{-g}\,(\sqrt{-\h{g}}$ yield for a variation 
of the charge an expression independent of $x$.  

In a general case the surface charges obtained by integration of variations of the 
charges over the space of metrics are path-dependent. 
In a special case of the supertranslation field in metric
depending only on a spherical angle $\th$, variation of the charge in the 
$r$-system is of the form of a variation  of a functional over the space of metrics
and can be integrated in a path-independent way. Integrability of the surface charge
was prouved for both parametrizations of the unit sphere.
Although of different functional forms, the expressions vor the variation of the charge
are transformed one to another by the transformation from the $r$ to the $\r$ system. 

\vspace{1cm}

{\large\bf Acknowledgments}

I thank Valery Tolstoy for useful conversation.

This work was partially supported by the Ministry of Science and
Higher Education of
Russian Federation under the Project 01201255504.


\section{Appendix A: Solution of Eq.(\ref{4.12})}
\setcounter{equation}{0}
\renewcommand{\theequation}{a\arabic{equation}}

In this Appendix we find a general solution for the function $T(z,\z )$
in the generator of supertranslations Eq. (\ref{4.1}). We solve the Eq. (\ref{4.12})
which is a condition on the generator $T$ at the horizon (written in 
notations of Sect.2)
\be
\l{a.1}
[-T\sqrt{1-b^2} +\f{1}{4}F^{ab}D_b T
(-2b_{a}\sqrt{1-b^2}+D_a b^2)]_{r=2}=0
.\ee
Eq. (\ref{a.1}) can be presented in a form
\be
\l{a.2}
T + F^a D_a T=0
,\ee
where
$$
F^a = \f{1}{2}F^{ac}(b_c + \pa_c \sqrt{1-b^2})|_{r=2}
$$
Following the general rules of
solving the differential equations with partial derivatives \cite{step},
we consider  a function
$W(T,z, \z )$ satisfying the equation
\be
\l{a.3}
T\f{\pa W}{\pa T} + F^z \f{\pa W}{\pa z} + F^{\z} \f{\pa W}{\pa \z }=0
.\ee
 Eq.(\ref{a.3}) is solved by writing  the system of ordinary differential equations
\be
\l{a.4}
\f{d T}{T} =\f{d z}{F^z}= \f{d \z}{F^{\z} }
.\ee
Let the independent first integrals of the Eq. (\ref{a.4}) be
\ba
\l{a.5}
\psi_1 (T,z,\z ) =C_1, \qquad \psi_2 (T,z,\z ) =C_2.
\ea
The general solution of the Eq. (\ref{a.3}) for  $W(T, z,\z )$ is
\be
\l{a.6}
W=f(\psi_1 ,\psi_2 ),
 \ee
where $f$ is an arbitrary smooth function.
The function $T(z,\z )$ is implicitrly determined from the equation
\be
f(\psi_1 ,\psi_2 )=0.
\ee

\section{Appendix B: Transformation $\not\d H_\xi (g,h)
 \longleftrightarrow\not\d H_\c (\h{g},\h{h})$ in the case $C=C(\th )$}
\setcounter{equation}{0}
\renewcommand{\theequation}{b\arabic{equation}}

In this Appendix we present details of transformations leading to the
relation  (\ref{6.m9}).

For $\r_s = 1/2 +x,\,\, |x|\ll 1$, using (\ref{2.4}) we have
\ba
\l{b.1}
&{}&\f{\pa r}{\pa \r}=\left(1-\f{1}{4\r_s^2}\right)\f{\pa\r_s}{\pa\r}=\f{(K^2-1)}{K^2}
\f{(\r -C)}{\r_s} \simeq 4(K-1)(\r -C),\\\l{b.2}
&{}&\f{\pa r}{\pa \th}=\left(1-\f{1}{4\r_s^2}\right)\f{\pa\r_s}{\pa\th}=\f{(K^2-1)}{K^2}
\f{( -C' )(\r -C -C'' )}{\r_s}\simeq 4( -C' )(\r -C -C'' )
.\ea
The derivative 
$\pa\r/\pa r$ with $\r$ (\ref{2.5}) is obtained as
\be
\l{b.3}
\f{\pa\r}{\pa r} =\f{K(\pa K/\pa r)} {4\sqrt{K^2/4 -(DC)^2}}\simeq [4(\r -C)(K-1)]^{-1}
.\ee
At the horizon,  $K=1$, Eqs.(\ref{b.1}) and (\ref{b.3})
give 
\be
\l{b.10}
\f{\pa r}{\pa\r}\f{\pa\r}{\pa r}=1.
\ee
In the same way, using (\ref{b.2}) and (\ref{b.3}), we obtain the second relation
(\ref{6.m8})
\be
\l{b.4}
\f{\pa \r}{\pa r}\f{\pa r}{\pa\th}=[4(\r -C)(K-1)]^{-1}
\f{(K^2-1)}{K^2}\f{( -C' )(\r -C -C'' )}{\r_s}
\simeq \f{( -C' )(\r -C -C'' )}{\r -C}=-\r _{,\th}.
\ee
Following transformations leading to (\ref{6.4}), we have
\be
\l{b.5}
\sqrt{-\h{g}}dr \w dt\w d\th\w d\p =\sqrt{-\h{g}_{\th\th}\h{g}_{\p\p}}
(\pa r/\pa\r)d\r \w dt\w d\th\w d\p .
\ee
To obtain (\ref{6.m9}), we use  the relation
\be
\l{b.6}
\h{g}_{\th\th}=\f{g_{\th\th}}{4(\r -C)^2}
\ee
which follows from the definition (\ref{2.111}).
We transform
\be
\l{b.7}
\sqrt{-\h{g}}dr =\sqrt{\h{g}_{\th\th}\h{g}_{\p\p}}(\pa r/\pa\r)d\r\simeq
\f{\sqrt{{g}_{\th\th}{g}_{\p\p}}}{2(\r -C)}4(K-1)(\r -C)d\r
.\ee
On the other hand,
\be
\l{b.8}
\sqrt{-g }d\r  \simeq \f{(\r_s -1/2)}{(\r_s +1/2)}4\sqrt{g_{\th\th}
g_{\p\p} }d\r  . 
\ee
Because $(\r_s -1/2)/(\r_s +1/2)\simeq (K-1)/2$, in the horizon limit $r\rightarrow 2$
expressions (\ref{b.7}) and  (\ref{b.8}) are equal..


\end{document}